# Integrating Intel SGX Remote Attestation with Transport Layer Security

Thomas Knauth, Michael Steiner, Somnath Chakrabarti, Li Lei, Cedric Xing, Mona Vij
Intel Labs
firstname.lastname@intel.com

## ABSTRACT

Intel® Software Guard Extensions (Intel® SGX) is a promising technology to enhance securely processing information in otherwise untrusted environments. An important aspect of Intel SGX is the ability to perform remote attestation to assess the endpoint's trustworthiness. Commonly, remote attestation is used to establish an attested secure channel to provision secrets to the enclave.

We integrate Intel SGX remote attestation with the establishment of a standard Transport Layer Security (TLS) connection. Remote attestation is performed during the connection setup by embedding the attestation evidence into the endpoints TLS certificate. Importantly, we neither change the TLS protocol, nor does our approach require changes to existing protocol implementations.

We have prototype implementations for three widely used open-source TLS libraries – OpenSSL, wolfSSL and mbedTLS. We describe the requirements, design and implementation details to integrate SGX remote attestation with TLS to bind attested TLS endpoints to Intel SGX enclaves.

## 1. INTRODUCTION

Intel SGX is a recent (2015) Intel processor extension available since the 6[th] Gen Intel® Core™ processors. Intel SGX enables application developers to construct trusted execution environments – called enclaves – to perform computation on commodity CPUs while achieving previously untenable security protections. Even highly privileged software running concurrently on the same hardware, say, the operating system, and virtual machine monitor, cannot observe an enclave's data in clear text. This is possible thanks to changes in the microarchitecture that enhances restriction-access to enclave memory by anyone except the enclave to which the memory belongs.

An integral part of the Intel SGX architecture is the ability to perform attestation. The *attester* wants to convince the *challenger* that it is a genuine Intel SGX enclave running on an up-to-date platform. At the end of the attestation process the enclave has convinced the challenger that it is genuine. Based on the enclave's attested attributes, the challenger decides whether to trust the enclave or not.

Previous work has shown that remote attestation and secure channel establishment must be integrated to protect against man-in-the-middle attacks [1]. At a minimum, the remote attestation protocol should result in a shared secret that can function as the basis for a secure channel. The current Intel SGX SDK provides an instance of this concept: remote attestation is performed using a modified Sigma [2] protocol. After a successful protocol instance the attester and challenger share a secret.

However, a shared secret only partially solves the problem of secure communication. Bootstrapping a secure channel based on a shared secret is possible, but inefficient since it duplicates work. Instead, we want to seamlessly integrate attestation with the establishment of a standard secure channel. The end result is an attested secure channel through which the participants can communicate.

In this white paper we describe our approach to combine Intel SGX remote attestation into the existing Transport Layer Security (TLS) secure channel protocol. Most importantly we do leave the TLS protocol unchanged, allowing us to reuse existing implementations. Additionally, our approach works with existing extension points in widely used and well-vetted implementations of TLS. Not requiring modified libraries and enabling developers already familiar with common secure channel API further, eases adoption. The proposed scheme is independent of the TLS implementation and we provide prototypes

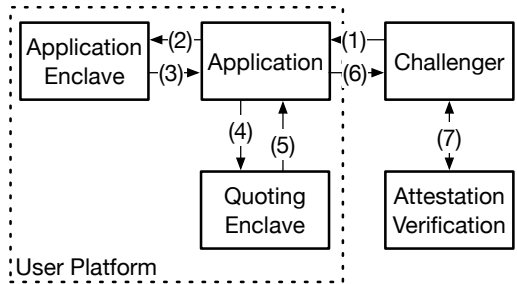

*Figure 1: Remote Attestation Example. The challenger is off-platform with respect to the attester.*

for three different popular TLS libraries – OpenSSL, wolfSSL and mbedtls.

The remainder of the text is structured as follows: We first provide some background on Intel SGX remote attestation and TLS in Section 2. Section 3 presents our approach to incorporate remote attestation into the TLS protocol. Section 4 gives implementation details and describes the API. In Section 5 we discuss future extensions and challenges. Section 6 discusses related work before concluding in Section 7.

## 2. BACKGROUND

In this section we provide background information on SGX local and remote attestation in general (Sec 2.1) followed by an overview of the EPID (Sec. 2.2) and ECDSA (Sec. 2.3) remote attestation model and the Transport Layer Security (TLS) protocol (Sec. 2.4). Readers familiar with either may want to skip ahead.

### 2.1. SGX ATTESTATION

The concept of attestation was previously explored in the context of Trusted Computing [3]. A secure co-processor, called Trusted Platform Module (TPM), performed cryptographic operations, such as key generation, signing and storing of keys, in a secure environment. In addition, a TPM can also attest the overall platform's state and configuration to other interested parties [4]. The assessing party is called *challenger* while the assessed party is the *attester*. Based on the attestation, the challenger can decide whether to trust the platform by comparing the attester's state to a reference value.

Attestation is also an integral feature of the Intel SGX architecture although the implementation details differ compared to TPMs. Intel SGX provides two attestation mechanisms [5] depending on where the challenger and attester reside. *Local attestation* allows two parties on the same platform to attest each other. The enclaves perform a local message exchange to assess their identities and the fact that they share a common platform. Note that local attestation alone is insufficient to establish whether the platform or the enclaves are genuine SGX enclaves. These facts can only be determined through *remote attestation* where the challenger typically is off-platform.

Intel SGX attestation covers the identity of the software running in the enclave (e.g., MRENCLAVE and MRSIGNER), non-measurable state, such as the enclave mode (e.g., debug vs. production), additional data the enclave wants to associate with itself (e.g., a manifest describing the software's configuration), and a cryptographic binding to the platform TCB. This information is bundled into a data structure called a report. An interested party inspects the attributes contained in the report to make a decision on the trustworthiness of the enclave.

The integrity and authenticity of the attestation report is ensured differently depending on whether the interested party resides on the same platform as the attester. For attestation between enclaves on the same platform, the report is authenticated through a report key. For remote attestation, a special quoting enclave signs the report with an attestation key to create a quote.

Figure 1 illustrates a high-level remote attestation flow [5] where the challenger and verifier engage in a modified Sigma protocol [2] to establish a shared secret. The protocol starts with the challenger sending a nonce to the application to guarantee freshness (Step 1). The application creates a manifest that includes a response to the challenge as well as an ephemeral key to encrypt future communication with the challenger (Steps 2 and 3). The application computes a hash of the manifest and includes this hash as user-defined data when creating the report. Including a hash of the manifest into the report is crucial, as this binds the ephemeral key to this enclave. Binding the key and enclave instance together is crucial to avoid masquerading attacks [6] [1] [7].

Next, the enclave generates a report that summarizes the enclave and platform state. The report includes information on the platform (security version number), enclave attributes, enclave measurement, software version, software vendor security version number, and additional user-provided data [8]. The quoting enclave verifies and signs the report using the attestation key (Step 4). The signed report, now called a quote, is returned to the application (Step 5) which passes it on to the challenger (Step 6).

The details of how the platform acquires the attestation key and how the quote can be verified (Step 7) differs between the two remote attestation models.

## 2.2. EPID ATTESTATION

Initial versions of Intel SGX focused on privacy-sensitive client platforms. The Enhanced Privacy ID (EPID) protocol [9] allows systems to be identified as genuine SGX platforms without revealing their identity in the process. This is important for use cases involving client platforms where the service provider wants assurance on the availability of SGX, but the end user wants to maintain privacy across attestations.

With EPID attestation, the platform's attestation key is established through a blinded join protocol run between the platform and an Intel back-end service [9]. As a result, the platform has a private group key, the attestation key, to sign attestations with. Anyone with access to the group public key can verify if the signature was generated by one of the many group private keys without learning which specific group private key generated the signature.

Instead of publishing the group public keys, Intel set up an Attestation Service to verify quotes. Within the EPID remote attestation model, a verifier sends the quote to the Intel Attestation Service (IAS) which reply with an *attestation verification report*, confirming or denying the authenticity of the quote and the enclave it originates from [10]. The verifier inspects the attestation verification report to determine the enclave and platform's trustworthiness. The reply from IAS includes the original quote and can thus be verified by 3rd parties other than the original challenger.

The Intel SGX SDK [11] provides the necessary APIs and primitives to remotely attest an enclave based on the modified Sigma protocol outlined earlier. Note though, that the Sigma protocol only results in a shared secret, not a secure channel. The application is left with the task to create a secure channel based on the shared secret. RA-TLS bridges this gap and integrates remote attestation into the establishment of the standard secure channel protocol TLS. This makes remote attestation easier to use securely in practice, reduces the barrier to entry and benefits the overall Intel SGX ecosystem.

## 2.3. ECDSA ATTESTATION

ECDSA-based attestation is an alternative attestation model that allows third parties to build their own non-Intel attestation infrastructure. ECDSA benefits environments, such as data centers, where EPID's privacy properties are unnecessary or deployment constraints, like the requirement to communicate with Intel back-end services during each attestation, make EPID infeasible. ECDSA-based attestation [12] relies on standard ECDSA signatures (hence the name) as opposed to anonymous group signatures as used in EPID. The quotes used to exchange information about an enclave and the platform carry an ECDSA signature to convey their authenticity and integrity.

With ECDSA-based attestation, Intel's active participation in the actual attestation flow is no longer required. Instead of going to the Intel Attestation Service to have a quote verified, a verifier can acquire all the necessary verification inputs without Intel's active participation including a particular certificate, issued by Intel, to assess an attestation. The Platform Certification Key (PCK) certificate can be obtained prior to an actual attestation taking place. Since the PCK certificate is valid for extended time periods (i.e., years) it can be cached and reused across many attestations.

While Intel provides a reference implementation for ECDSA-based attestation along with a software library to generate and verify quotes [13], third parties are free to set up their own attestation infrastructure, including writing their own quoting enclave.

## 2.4. TRANSPORT LAYER SECURITY

Transport Layer Security (TLS) is the de-facto industry standard to secure communication. It has been developed since 1995 and is still evolving, with TLS version 1.3 being the latest revision. Even with this long history, subtle deficiencies are still discovered occasionally [14]. Thus our strong motivation to build on a standard secure channel protocol instead of custom solutions.

**Authentication.** TLS not only protects the integrity and confidentiality of data but also allows endpoint authentication through the exchange of X.509 certificates. In a typical scenario where a browser (client) connects to a web site (server), only the server sends a certificate. The client uses the certificate to confirm that it indeed connected to the intended web site.

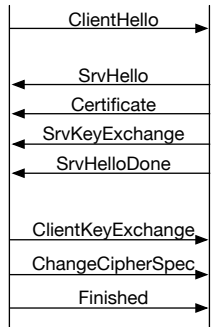

*Figure 2: TLS 1.2 Handshake Messages.*

TLS also allows both endpoints to authenticate each other. Mutual authentication is frequently used with known client populations, for example, in enterprise settings, where only previously registered clients are allowed to connect. Again, trust in the endpoint is conveyed through the exchange of certificates and their verification through a certificate chain linking back to a trusted root certificate.

**X.509 Certificates.** The use of certificates for the purpose of identification is based on a Public Key Infrastructure (PKI). A set of trusted root certificate authorities either create leaf certificates directly or delegate the responsibility to an intermediate certificate authority (CA). The intermediate CA can issue leaf certificates "on behalf of" the root CA. In the case of securing web sites with TLS, the certificate represents a cryptographic binding of the web sites domain name to the public key (subject key) referenced in the certificate. When connecting to the HTTPS web site, the client verifies the certificate chain as well as that the distinguished name in the certificate matches the domain it intended to connect to.

Besides the distinguished name, X.509 certificates can be used to bind arbitrary data to the key identity in the form of X.509 extensions. Each extension is identified by an Object Identifier (OID). For example, the OID for the well-known extension Subject Alternative Name (SAN) is 2.5.29.17. SAN allows the certificate to list multiple domain names for which the certificate is valid. X.509 extensions are not regulated. Companies are free to introduce new extensions in combination with the products they offer. To ensure interoperability, extensions can be marked critical. If a client encounters a critical extension it is unaware of, it aborts the handshake. A client is, however, free to ignore an unknown non-critical extension.

**Trust Root.** With TLS and the use of X.509 certificates to identify the endpoint, the trust root lies in the root certificate authorities. PKI users trust the root CAs to follow best practices and procedures before issuing certificates to service providers. Alas, not all CAs and their delegates are created equal and some may be more trustworthy than others [15]. While browsers list hundreds of trusted CAs, scenarios other than HTTPS on the internet may use a much smaller set of trusted roots.

**TLS Handshake.** At the beginning of a TLS connection, the client and server engage in a 3-way handshake[1]. As part of the handshake, the two parties agree on the specifics of the cipher suite and the server authenticates itself to the client by sending a certificate. While the client initially encrypts messages to the server using the server's public key obtained from the certificate, the two parties establish a mutual session key to use after the handshake.

 illustrates a typical TLS 1.2 handshake. The client initiates the handshake by sending a ClientHello message. The ClientHello includes a nonce, specifies the client's preferred cipher suites and any supported TLS extensions. The server replies with a sequence of three messages: ServerHello, Certificate, and KeyExchange. After the server sends its ServerHelloDone message, the client continues the exchange with its KeyExchange and ChangeCipherSpec messages. After the handshake finishes successfully, the parties will have established a session key to authenticate and en-/decrypt future messages. A detailed description of each message and their meaning is available in the official TLS 1.2 standard [16]. TLS 1.3 coalesced some previously separate handshake messages to reduce the number of round trips. Otherwise, the changes TLS 1.3 introduces do not affect how it integrates with Intel SGX remote attestation.

---

[1]We are focusing on TLS v1.2 here.

As part of the handshake, the server sends an X.509 certificate to the client. The certificate states the server's public key, the certificate's validity period, and domain name(s) the certificate is valid for, among other things. The client verifies this information and performs path validation to ensure the certificate chain terminates in a trusted root certificate. Additionally, the client should also check if the certificate has been revoked. In practice, this is only done for Extended Validation (EV) certificates, a special class of certificates used by high-value web sites such as banks and insurance companies. If the client's certificate validation logic indicates a problem with the certificate, the handshake is terminated.

During the handshake, the server and client negotiate a symmetric **session key** to use after completing the handshake. The specifics of the key exchange and the authentication depend on the chosen cipher suite. For example, a cipher suite employing the Diffie-Hellman Ephemeral key exchange, results in a new key for each session. A unique session key provides forward secrecy, that is even if the server's key is compromised, an attacker cannot decrypt previously recorded sessions since each session used an ephemeral key independent of the server's key.

## 3. DESIGN

In contrast to the PKI used in the modern web where the trust root is a list of root certificate authorities maintained by a handful of entities such as Google, Mozilla and Microsoft, we want to use Intel SGX as a hardware root of trust. To this end, we propose to include additional information into the X.509 certificate exchanged during a TLS handshake. In the following, we explain which information we include in the certificate and how this information is validated.

We stay with the common scenario where a central server allows arbitrary clients to connect and the clients want to verify the server's identity during the connection setup. In the terminology of remote attestation, the server is the attester and the client the challenger. We focus on the classic client/server scenario to ease the explanation but our scheme also covers mutual authentication/attestation. With mutual attestation, the client and server assume the roles of attester and challenger at the same time.

The server runs in an Intel SGX enclave and terminates the TLS connection inside the enclave. The server can either be constructed using the Intel SGX SDK or any other framework/runtime to execute legacy applications on Intel SGX [17] [18] [19] [20]. The server obtains its Intel SGX identity and executes the steps necessary to obtain a report and a quote from the platform's quoting enclave. This is required to successfully complete the remote attestation workflow.

To achieve our goal of integrating remote attestation into the TLS handshake, we have to address two objectives: First, we need to link the server's RA-TLS key to a specific instance of an Intel SGX enclave. Second, the server/attester must provide the necessary attestation evidence during the TLS connection setup to convince the client/challenger that it is indeed connected to a veritable Intel SGX enclave.

### 3.1. BINDING RA-TLS KEY TO ENCLAVE

The enclave generates a new public-private RA-TLS key pair at every startup. The RA-TLS key need not be persisted since generating a fresh key on startup is reasonably cheap. Not persisting the key reduces the key's exposure and avoids common problems related to persistence such as state rollback protection. Interested parties can inspect the source code to convince themselves that the key is never exposed outside of the enclave.

RA-TLS links the RA-TLS key and enclave by including a hash of the RA-TLS public key as user-data into the Intel SGX report. Recall, that the report is generated by the attester and passed to the platform's quoting enclave. The quoting enclave signs the report, vouching that the report was indeed generated by a genuine Intel SGX enclave on the local platform. At this point, there exists a cryptographically secured statement of the fact that the RA-TLS public key is bound to the Intel SGX enclave.IAS Intel

### 3.2. ATTESTATION EVIDENCE

To assess the server's Intel SGX identity and link its public key to the enclave instance, the client needs access to the attestation evidence. We propose to embed the attestation evidence as custom X.509 extensions in the server's certificate. To this end, we introduce new X.509 extensions, each with their own unique object identifier (OID). Adding the attestation evidence into the certificate is less intrusive than, for example, introducing a new TLS extension which would require invasive changes to each TLS implementation. By extending the certificate, we do not require any changes to existing TLS libraries and use readily available hooks to verify the custom X.509 extensions.

Extending the certificate traditionally requires resigning it by a CA. However, since we propose to use Intel SGX as a trust root, we can simply self-sign the certificate. Instead of relying on the CA to bind the domain name to the server's identity/key, we rely onIntel SGX to provide the identity. If a binding between the Intel SGX identity and a domain name is desired, we discuss one proposal how to incorporate this into the attestation flow in Section 5.

The specific attestation evidence embedded into the RA-TLS certificate differs between EPID and ECDSA-based attestation as is described next.

### 3.2.1. EPID ATTESTATION EVIDENCE

For EPID-based attestation [9], we propose to embed the following items as attestation evidence into the RA-TLS certificate.

**Attestation Verification Report.** This is the reply received from the Intel Attestation Service (IAS) when submitting a quote for verification [10]. IAS reports whether the quote was generated on a genuine Intel SGX platform by a genuine enclave. Embedded in the report is a copy of the data produced by the enclave, such as platform security version number, enclave identity (MRENCLAVE) and enclave signer (MRSIGNER) among other attributes.

**Attestation Verification Report Signature.** The attestation verification report is signed by IAS. The signature ensures that report is authentic and unmodified.

**Attestation Report Signing Certificate.** To verify the signature over the attestation report, the client must know the keys used to create the signature. This information is conveyed in additional certificates included in the response from IAS.

In addition to the above items the verifier also needs the Attestation Report Signing CA Certificate to establish whether the Attestation Report Signing Certificate is trusted. This certificate forms the trust root and is thus not included in the RA-TLS certificate. We expect the verifier to obtain it through a secure out-of-band mechanism.

Also note that in RA-TLS the attester communicates with IAS since it assembles all the attestation evidence to be consumed by the verifier. This is different to the flow depicted earlier in Figure 1 where the verifier only receives the quote and handles the verification independently. Depending on the scenario, it may be important which party has to communicate with other external services during the attestation flow.

### 3.2.2. ECDSA ATTESTATION EVIDENCE

For ECDSA-based attestation, we propose to include the following items as attestation evidence into the RA-TLS certificate.

**Quote.** The signed data structure produced by the quoting enclave. It contains the enclave's code identity and various attributes. The quote is signed with the attestation key. The verifier can examine the quote to determine whether to trust the enclave or not based on its SGX identity.

**TCB Info.** Collection of attributes on the platform's state where the quote was generated. A verifier can examine this to determine whether or not to trust the platform on which the attestation was performed. For example, a verifier may decide to dismiss any attestation originating on a platform missing critical security updates.

**TCB Signing Chain.** The TCB info is signed to assess authenticity and integrity. The certificate chain required to verify the signature is included as part of the attestation evidence.

**PCK Certificate.** Intel publishes a PCK certificate for each SGX platform and TCB. The PCK certificate allows a verifier to determine if the quote was produced on a genuine SGX platform.

**PCK Signing Chain.** The PCK certificate can be traced back to a trusted root certificate owned by Intel. The attestation evidence includes the entire certificate chain. The verifier is expected to check the signature on the PCK certificate to ascertain authenticity.

**Certificate Revocation Lists.** For the PCK certificates and intermediate CA certificates Intel also publishes certificate revocation lists (CRLs) to notify a verifier of compromised keys.

**Quoting Enclave Identity.** The latest identity of Intel's quoting enclave including its code measurement and security versions numbers. This allows the verifier to assess if the quoting enclave is from a trusted source and received adequate security patches.

The TCB Info, PCK Certificate and Certificate Revocation Lists are all signed and chain back to a trusted root key possessed by Intel. The verifier is expected to have the root key's certificate (or the expected public key) either embedded or to otherwise obtain it securely out-of-band.

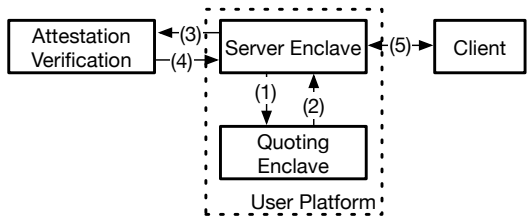

*Figure 3: Message flow for EPID-based remote attestation using TLS.*

### 3.3. RA-TLS CERTIFICATE VALIDATION

The receiver of the RA-TLS certificate must verify the SGX attestation evidence contained within. The concrete steps differ based on the attestation method. Protocol-wise, the verification happens during the standard TLS handshake. If any of the verification steps fail, the endpoint aborts the handshake to terminate the TLS connection.

**EPID Attestation Verification.** To verify the EPID attestation evidence a challenger must complete the following steps: (1) Verify the signature on the Attestation Report Signing Certificate against the Attestation Report Signing CA Certificate (which is trusted). Only if validation passes do we have assurance that the report was signed by a trusted attestation service. (2) Verify the report signature against the report using the Attestation Report Signing Certificate to ascertain that the attestation report is genuine and unmodified.

**ECDSA-based Attestation Verification.** In general, a verifier must undertake the following steps to determine the trustworthiness of an enclave on a particular platform [12]: (1) Verify the integrity of the signature chain from the quote to the Intel-issued PCK certificate. (2) Verify no keys in the chain have been revoked. (3) Verify the quoting enclave is from a suitable source and up-to-date. (4) Verify the status of the Intel SGX TCB described in the chain.

**RA-TLS Verification.** Independent of whether EPID or ECDSA attestation is used, the verifier must check if the hash of the RA-TLS certificate's public key is present in the enclave's report. This is how the RA-TLS key is tied to a particular enclave instance. Finally, the verifier compares the enclave's identity (e.g., MRENCLAVE and MRSIGNER) against the expected identity.

### 3.4. SECURITY PROPERTIES

Typically, SGX attestations are produced at time of use, guaranteeing a fresh attestation. For RA-TLS, since the attestation is produced when the key is generated, additional mechanisms are needed.

First, freshness of the exchanged messages is guaranteed by TLS itself: nonces at the record layer protect against replays. Freshness of the attestation evidence can be determined through time stamps attached to it. With EPID-based attestation, the attestation verification report has a time stamp. Similarly, with ECDSA-based attestation, the TCB info and certificate revocation lists have a validity period. Based on the available time stamps a verifier might reject them as too old.

The security of the RA-TLS key is ensured by Intel SGX. The quote binds the key to a particular enclave instance. By design, only the enclave has access to the key. An attacker would need to break the Intel SGX security model or exploit application-specific vulnerabilities to compromise the key. Users can inspect the source for code paths that intentionally expose the key. We do not intend to persist the key across enclave restarts. Instead, a new key is generated at each startup and, possibly, periodically at runtime for long running services. Usage of the key is further minimized when a cipher suite with a Diffie-Hellman Ephemeral key exchange is used. In this case, the RA-TLS key is only used during the handshake and a per-connection ephemeral Diffie-Hellman key protects the subsequent communication.

A classic problem with traditional PKI is the regeneration of certificates, either because the certificate expired or the key was compromised, which is time-consuming and expensive. Since we root trust in the Intel SGX hardware, we can self-sign the certificate. This allows us to have short expiration times and regenerate certificates periodically. Whenever we change the RA-TLS key, we do need to generate a new quote to reflect this. Short certificate validity periods and fresh attestations also protect challengers from connecting to revoked platforms. For example, with EPID attestation, the attester will no longer be able to obtain a valid attestation report from the Intel Attestation Service if a platform was revoked in the meantime.

```
01 WOLFSSL_CTX* ctx;
02 ...
03 char key[2048]; char crt[8192];
04 int key_len = sizeof(key);
05 int crt_len = sizeof(crt);
06
07 create_key_and_x509(key, &key_len,
08                     crt, &crt_len);
09
10 wolfSSL_CTX_use_certificate_buffer(ctx,
11     crt, crt_len, SSL_FILETYPE_ASN1);
12
13 wolfSSL_CTX_use_PrivateKey_buffer(ctx, der_key,
14     der_key_len, SSL_FILETYPE_ASN1);
```

*Figure 4: An example, based on the wolfSSL library, using the attester API. Functions exported by the remote attestation library are in bold.*

For ECDSA-based attestation, if the verifier is concerned about the attestation evidence's freshness, it is possible to independently obtain some of it instead of relying on the material contained within the RA-TLS certificate. This applies in particular to certificate revocation lists to ensure none of the keys used to sign attestation evidence or the PCK key have been revoked. Especially for verifiers that themselves run inside an enclave it may be difficult to determine if the attestation evidence is past its validity period without access to a trusted time source.

## 4. IMPLEMENTATION

We implemented RA-TLS for EPID-based attestation using three different TLS libraries: wolfSSL [21], mbedtls [22], and OpenSSL [23]. ECDSA-based attestation in RA-TLS is currently only implemented in wolfSSL[2]. Overall, the implementation is structured into two main components: (1) the RA-TLS key and certificate generations for the attester. (2) The RA-TLS certificate validation for the challenger. The RA-TLS C library exposes a set of functions for use by the attester and challenger. The library is linked into the application like any other standard library.

The code clearly separates the TLS library-specific parts (hash computation, certificate generation and validation, and key generation) from the library agnostic aspects (extended validation steps). This allows us to reuse the library-agnostic parts when porting RA-TLS to other TLS libraries. The interface to the RA-TLS library is largely independent of the attestation model expect for a structure to pass in configuration parameters. For example, EPID requires a Software Product ID (SPID) while ECDSA requires an API subscription key to talk to the respective backend services.

Besides the actual TLS library, we also encapsulate the Intel SGX SDK specific functionality into separate functions. For example, the RA-TLS library uses the Intel SGX SDK to obtain the quote. If the Intel SGX SDK is not available, for example, because a different framework is used [19] [18] [20] to execute the program on Intel SGX hardware, we provide an alternative implementation of the required functionality independent of the Intel SGX SDK. Hence, integrating our library with other SGX frameworks is possible and low overhead.

Next, we describe the programming interface in terms of the attester and the challenger. Depending on the scenario, either can be the server or client in a traditional distributed computing setting. If mutual attestation is desired, and this is supported by our library and TLS, the client and server will in fact assume the role of attester and challenger at the same time.

### 4.1. ATTESTER

Traditionally, the server/attester reads the key and certificate from stable storage into memory before serving it to other parties. It is the attester's responsibility to apply sufficient security measures to safeguard the key at rest.

Instead we propose to generate a new key and certificate whenever the attester enclave starts. The key and certificate are ephemeral and shall not be persisted past the attester's lifetime. The attester's API consists of one function `create_key_and_certificate` that outputs the key and the corresponding certificate. To keep the interface generic, the key and certificate are encoded in standard DER format.

 illustrates the flow of messages sent by the server when it initializes. The server runs inside an Intel SGX enclave. Before accepting any client connections, the server generates a new key and creates a self-signed X.509 certificate including the extensions mentioned in the previous section. The TLS library is configured to use this certificate as the server certificate for new client connections.

---

[2] RA-TLS uses Intel's Data Center Attestation Primitives [13] for the actual verification of the attestation evidence which uses OpenSSL as a backend.

```
01 int cert_verify_cb(int preverify,
02   WOLFSSL_X509_STORE_CTX* store) {
03
04   WOLFSSL_BUFFER_INFO* crt = store->certs;
05   int ret = verify_sgx_cert_extensions(
06                  crt->buffer,
07                  crt->length);
08   return !ret;
09 }
10
11 int main(int argc, char* argv[]) {
12   ...
13   wolfSSL_CTX_set_verify(ctx, SSL_VERIFY_PEER,
14                  cert_verify_callback);
15   ...
16   WOLFSSL_X509* crt =
17       wolfSSL_get_peer_certificate(ssl);
18
19   int der_len;
20   const byte* der =
21       wolfSSL_X509_get_der(crt, &der_len);
22
23   sgx_quote_t quote;
24   get_quote_from_cert(der, der_len, "e);
25   sgx_report_body_t* body = "e.report_body;
26
27   if (0 != memcmp(body->mr_enclave.m,
28                   golden_mr_enclave,
29                   SGX_HASH_SIZE)) { exit(1) };
30   ...
31 }
```

*Figure 5: An example, based on wolfSSL library, using the challenger API. Functions exported by the remote attestation library are in bold.*

To create the certificate, the server must go through the usual steps to perform remote attestation: create a report, pass the report to the quoting enclave, and, finally, send the quote to IAS to receive an attestation verification report. These steps must be performed at startup (and every time the server's key changes at runtime). If it is desirable to maintain the same key across server restarts, the key can be sealed to the enclave's identity.

We illustrate the usage of the attester API in . In this particular example we use the wolfSSL library to configure an endpoint accepting connections. After defining the data structures to hold the key and certificate (lines 3–5) we call our library function to create a new key and certificate (lines 7–8). Subsequently the attester calls the wolfSSL function to use the newly generated key (lines 10–11) and certificate (lines 13–14) with incoming connection requests. The simple and intuitive API enables developers to integrate our library easily into existing applications.

### 4.2. CHALLENGER

The challenger extends the existing certificate validation logic. To determine if a server certificate is valid, the TLS library performs a standard set of checks: it verifies the certificate's expiration status, if the certificate's content matches its signature and whether the certification chain terminates in a trusted root certificate.

All the TLS libraries we have examined allow the user to specify a custom certificate validation function. The custom validation function either extends the built-in verification logic or overrides it depending on how it is used. Using these hooks/callbacks, we implement all the required changes to the challenger without modifying the TLS library itself.

The certificate validation hook indicates the successful validation through the return value. The caller uses the hook's return value to either continue the handshake or abort it. The hook calls the function `verify_sgx_cert_extensions` exported by our library. If the verification succeeds, the function returns 0. Otherwise, the function returns 1.

The application gets access to the Intel SGX identity and platform attributes through a complimentary function `get_quote_from_cert` to extract the Intel SGX quote from the X.509 certificate. The quote contains information on Intel SGX identity attributes such as MRENCLAVE and MRSIGNER as well as platform related attributes such as the CPU security version number. Challengers use the attributes to decide whether the remote end meets their security requirements. It is also the application which must perform authorization based on the SGX identity while the RA-TLS library performs the authentication.

 illustrates how to use the challenger API in a wolfSSL-based application. First, we define a custom certificate verification function `cert_verify_cb` with the signature expected by the wolfSSL library (lines 1–9). This verification callback simply wraps our library function doing the extended certificate validation. Later on, we register the certificate verification callback with the wolfSSL library (lines 13–14). After the TLS handshake finished successfully, we obtain access to the peer's DER encoded certificate (lines 16–21). Next, the challenger extracts the quote from the DER encoded certificate (lines 23–25). The peer's MRENCLAVE value is compared against a reference value to verify the SGX identity (lines 27–29).

Even though the example is based on wolfSSL, the flow is similar for other TLS libraries. The data structures and function signature will be different, but the interface exposed by our library is generic to make interoperability with other TLS libraries easy.

## 5. LIMITATIONS AND EXTENSIONS

We discuss limitations and considerations when pursuing seamless attestation via extended X.509 certificates.

**Non-standard X.509 extensions.** We embed the SGX identity information in custom X.509 extensions. X.509 extensions can be marked as critical. If a client encounters an unknown critical extension, the default is to abort the connection. If the attester only expects Intel SGX-aware clients, it is sensible to mark the extensions critical. If the client includes Intel SGX-unaware legacy clients, the extensions should not be marked critical to allow backwards compatibility. A legacy client may still complain about a self-signed certificate.

Clients wishing to make use of the Intel SGX identity included in the X.509 certificate need to extract this data. To reuse existing APIs, it may make sense to fold the Intel SGX identity attributes into established X.509 fields. For example, the Common Name (CN) attribute could hold a copy of MRENCLAVE. When validating the certificate's Intel SGX extensions fields, the verifier must verify that the Intel SGX values encoded in the common attributes match the ones embedded in the quote.

**Certificate size.** Including all the attestation evidence into the RA-TLS certificate increases its size significantly compared to established use cases. For example, a typical HTTPS certificate is around 1100 byte (DER format)[3] while an RA-TLS certificate is around 6200 bytes for EPID attestation and at least 11,000 bytes for ECDSA attestation.

The size increase of the certificate may be relevant for attesters that establish new connections frequently since the certificate's size dominates the overall handshake-related traffic volume. Also, some TLS implementations use statically allocated buffers that might be too small to hold a large certificate leading to failed validations.

One option is to compress the attestation evidence stored in the certificate. A lightweight and fast compression algorithm such as LZ can already reduce the certificate's size by about 50%. If the client population is stable, TLS session resumption reduces the number of times the certificate must be transmitted.

A significant portion of the attestation evidence consists of infrequently changing certificate chains and revocation lists. While it is certainly convenient to have all the attestation evidence available from a single source, it may be more cost effective if the verifier obtains these long-lived inputs through a different channel and caches them. As a result, the attester does not need to transmit the long-lived evidence with every TLS handshake.

**Self-signed certificate.** The attester creates a self-signed certificate that represents its identity. This may pose problems with legacy clients. For example, in the case of HTTPS, clients expect a known trusted CA to sign the certificate. If the certificate is self-signed, the client's certificate validation logic will abort the handshake. A self-signed certificate will also lose the binding between the server's domain name and the key as vouched for by the CA.

A solution around this dilemma is to have a trusted CA sign the extended certificate. With protocols such as the Automated Certificate Management Environment (ACME) to streamline the issuance of certificates [24], this is one possibility to combine traditional X.509 identities with Intel SGX identities. Alternatively, if the attester has access to an intermediate CA certificate, this can be used to sign the Intel SGX-extended certificate. In this way, the leaf certificate chains back to a PKI trust root, albeit at the added burden to securely handle the intermediate CA's key.

**Certificate Revocation Lists (CRLs).** For ECDSA, the verifier does not know if TCB information or PCK CRLs are current. An attacker may break a platform and use outdated CRLs to convince a verifier of its trustworthiness. A non-enclave verifier can inspect the CRLs validity period to determine if they are outdated. Due to the lack of a trusted time source this is impossible for a verifier running in an enclave.

To protect against outdated CRLs and certificates, instead of relying on CRLs contained in the RA-TLS certificate, the verifier may want to obtain CRLs independently at appropriate intervals. Alternatively, it might make sense to introduce an OCSP service or OCSP-staple responses of the API endpoints serving the TCB info and PCK certificates. Ultimately though, even the stapled versions of PCK certificates and TCB information may be outdated and assessing the freshness of the attestation evidence requires a trusted time source within the enclave.

---

[3] https://scans.io/data/rapid7/sonar.ssl/20170912/20 17-09-12-1505178001-https_get_443_certs.gz -- We used this data set for the certificate related statistics.

## 6. RELATED WORK

The TLS protocol provides authentication, integrity, and privacy for data transmitted across untrusted communication lines. However, people have cautioned that "Using encryption on the Internet is the equivalent of arranging an armored car to deliver credit-card information from someone living in a cardboard box to someone living on a park bench." [25]. The reason being that it is far easier to subvert the endpoint than the channel.

Leveraging novel hardware security features, like Intel SGX, makes it significantly harder to subvert the endpoint. Still, moving the endpoint into an enclave is insufficient. What is needed is a way to bind the endpoint to a particular execution context, i.e., an enclave. Otherwise, the system is susceptible to relay attacks, where a compromised endpoint claims to have certain properties, but the properties actually pertain to a third system also controlled by the attacker.

Previously work in this direction [7] falls under the general umbrella of Trusted Computing [3] and focused on Trusted Platform Modules as the ultimate trust anchor. Intel SGX enables a more performant and versatile solution compared to a resource constrained TPM. Hence, it is worth revisiting the problem of secure channel binding in the context of Intel SGX.

In [6] the authors introduce an attestation extension to TLS. The peers exchange attestation information during the handshake. They used provisions within TLS to incorporate the additional information [26]. This particular TLS extension seems to be superseded by HelloExtensions messages in more recent versions of TLS. We decided to embed the additional channel binding information into the certificate to minimize changes to the TLS library. Introducing new messages [27], would have required changes to the library itself. Extending the certificate validation logic is achieved by calling TLS library hook functions from the application. It does not require any changes to the TLS library, making it easier to maintain and port to new TLS libraries. Complex interactions between existing extensions already plague the overall security of TLS [28].

Armknecht et al. [1] also identified the channel binding problem as crucial to the value remote attestation provides. Armknecht et al. propose a generic protocol that combines key exchange and remote attestation to avoid relay/masquerading attacks. However, they do not integrate the generic protocol into any existing secure communication protocol like TLS or IPsec. We integrated our solution with three existing open-source TLS libraries, making it directly usable to applications building on those libraries.

In subsequent work [29], Armknecht et al. propose alternative protocols to address performance concerns with software that frequently needs to provide attestations. Since our work is based on Intel SGX, as opposed to a TPM, these performance concerns do not apply. Attestation is cheap and its overall cost dominated by the communication latency to IAS. Since challengers do not contact the attestation service there is no attestation-related communication overhead added to the TLS handshake. The information required to assess the channel binding properties are embedded in the certificate and the challenger can verify them locally.

Most similar to RA-TLS is the Finite State Attestation Protocol (FSA) proposed by Paverd [30]. While FSA uses TPM attestations, it also proposes to bind the secure channel to the attestation by embedding a fingerprint of the ephemeral TLS key into the attestation.

## 7. CONCLUSIONS

Intel SGX offers a unique opportunity to enhance secure computation in otherwise untrusted environments. An integral part of Intel SGX is the ability to obtain an attestation on the properties of the enclave and its platform. Integrating remote attestation seamlessly with a standard secure channel protocol greatly simplifies the use of remote attestation in practice. We developed a library that conveniently encapsulates the attestation flow and verification behind a simple API. Using this interface, developers can rely on the added assurance remote attestation provides their application without having to deal with the intricacies of implementing it correctly.

A proof of concept implementation of remote attestation integrated into the TLS handshake is available at https://github.com/cloud-security-research/sgx-ra-tls

*Acknowledgements*

The authors thank Vincent R. Scarlata and Daniel E. Smith for comments on earlier drafts of this paper.